\documentclass[12pt]{article}

\usepackage{a4}
\usepackage{epsfig}

\newlength{\abstwidth}
\begin{document}

\def\lsim{\mathrel{\rlap{\lower4pt\hbox{\hskip1pt$\sim$}}
    \raise1pt\hbox{$<$}}}         
\def\gsim{\mathrel{\rlap{\lower4pt\hbox{\hskip1pt$\sim$}}
    \raise1pt\hbox{$>$}}}         

\pagestyle{empty}

\begin{flushright}
\end{flushright}

\vspace{\fill}

\begin{center}
{\Large\bf Low-x scaling\\[1ex]
in $\gamma^*p$ total cross sections${}^{*}$}\\[1.8ex]
{\bf Dieter Schildknecht} \\[1.2mm]  
Fakult\"{a}t f\"{u}r Physik, Universit\"{a}t Bielefeld \\[1.2mm] 
D-33501 Bielefeld \\[1.5ex]
{\bf Bernd Surrow} \\[1.2mm]
Max-Planck Institut f\"{u}r Physik (Werner-Heisenberg-Institut) \\[1.2mm]
F\"{o}hringer Ring 6, D-80805 M\"{u}nchen \\[1.5ex]
and \\[1.5ex]
{\bf Mikhail Tentyukov${}^{**}$} \\[1.2mm]
Fakult\"{a}t f\"{u}r Physik, Universit\"{a}t Bielefeld \\[1.2mm]
D-33501 Bielefeld 

\end{center}

\vspace{\fill}
\begin{center}
{\bf Abstract}\\[2ex]
\begin{minipage}{\abstwidth}

We show that the experimental data for the total virtual-photon proton
cross section, $\sigma_{\gamma^*p} (W^2, Q^2)$, for $x_{bj} \lsim 0.1$
lie on a universal curve, when plotted against $\eta = (Q^2 + m^2_0)/
\Lambda^2(W^2)$, where $\Lambda^2 (W^2) = C_1 (W^2+W^2_0)^{C_2}$ is
determined by the parameters $C_1, C_2$ and $W^2_0$. The observed
scaling law follows from the generalized-vector-dominance/colour-dipole
picture (GVD/CDP) of low-x deep inelastic scattering.

\end{minipage}
\end{center}

\vspace{\fill}
\noindent


\vspace{\fill}
\noindent

\rule{60mm}{0.4mm}

\vspace{0.1mm}
\noindent
${}^*$ Supported by the BMBF, Bonn, Germany, Contract 05 HT9PBA2\\
${}^{**}$ On leave from BLTP JINR, Dubna, Russia 
\clearpage
\pagestyle{plain}
\setcounter{page}{1}

The present note will be concerned with deep inelastic scattering (DIS) in the
kinematic range of low $x_{\rm bj} \simeq Q^{2}/W^{2} \ll 0.1$ that has been
and is being explored at HERA. In particular, we will show that the 
data \cite{1,2,3,4,5,5a} 
on photo- and electroproduction,
for
$x<0.1$, in good approximation, lie on a single curve, when plotted against 
the
dimensionless (`low-$x$ scaling') variable

\begin{equation}
\eta=\frac{Q^{2}+m_{0}^{2}}{\Lambda^{2}(W^{2})}.
\end{equation}
The value of the threshold mass, $m_{0}<m_{\rho^{0}}$, as well as the 
parameters
$C_{1}$, $C_{2}$ and $W_{0}^{2}$ contained in
\begin{equation}
\Lambda^{2}(W^{2})=C_{1}(W^{2}+W_{0}^{2})^{C_{2}}
\end{equation}
are fixed by the experimental data themselves.

We will proceed in two steps, the first one being a purely empirical analysis
of the data, while
in the second step, we will show, how the observed behaviour of
$\sigma_{\gamma^* p}(W^2, Q^2)$ can be understood in terms of generalized
vector dominance (GVD) \cite{6,7}\footnote{Compare also ref.\cite{7a} for
photo-and electroproduction off nuclei.}
or, equivalently \cite{8}, \cite{8a} 
the colour-dipole picture
(CDP) \cite{9}. 
\begin{enumerate}
\item[i)]
In the first step, the model-independent phenomenological analysis of the 
experimental data, we assume the analytic form of the scaling variable $\eta$
according to (1) and (2), and, in addition, the existence of a continuous 
function of $\eta$
that is supposed to describe the data for $\sigma_{\gamma^* p} (W^2,Q^2)$, 
when these are plotted
against $\eta$. For the technical analysis, we assume that this continuous
function,
without much loss of generality, may be represented by a piecewise linear
function of $\eta$. This assumption allows us to
perform a fit that determines the values of the 
parameters\footnote{
  In the model-independent fit, the parameter $C_1>0$ is an arbitrary input 
parameter as with a scaling function $f(\eta)$ also $f(C_1^{-1}\eta)$
describes  the data. The model-independent fit was carried out for several
values of $C_1$ around $C_1=0.34$ as uniquely determined in  the fit based on the
GVD/CDP to be described below.
} $m_{0}^{2}$,
$C_{2}$ and $W_{0}^{2}$ 
simultaneously with the values
of the piecewise linear function $\sigma_{\gamma^{*} p}(\eta )$ at a
number of points, $\eta_i (i=1,...,N)$, of the variable $\eta$.
\item[ii)]
In the second step, we show how an approximate scaling law in terms of the
variable $\eta$ follows from GVD or the CDP. We will restrict ourselves to 
only
present the essential theoretical assumptions and conclusions. For a detailed
account, we have to refer to a forthcoming paper \cite{10}.
\end{enumerate}

Turning to step i), in fig.1, we show the result of the 
model-independent analysis. Imposing the kinematic restrictions of 
$x \leq 0.1$ and $Q^{2} \leq 1000\,$GeV$^{2}$, all available experimental 
data \cite{1,2,3,4,5,5a} on photo- and electroproduction
are indeed seen to lie on a smooth curve, that is
approximated by the piecewise linear fit curve. The parameters 
that determine the scaling variable $\eta$ were found to be given by
\begin{eqnarray}
m_{0}^{2} & = & 0.125 \pm 0.0.027\; {\rm GeV^{2}}, \nonumber \\
C_2       & = & 0.28 \pm 0.06, \nonumber \\
W_{0}^{2} & = & 439 \pm 94\; {\rm GeV^{2}}, 
\end{eqnarray}
with a $\chi^{2}$ per degree of freedom ($ndf$) of $\chi^2/ndf=1.15$.

We add the remark, that an analogous procedure applied to the experimental 
data,
without a restriction on $x$, does {\it not} lead to a universal curve.
Likewise,
restricting oneself to only those data points that belong to $x>0.1$, {\it no}
universal curve is obtained either; the fitting procedure leads to entirely
unacceptable results on the quality of the fit as quantified by the 
value of $\chi^{2}$ per
degree of freedom.

We turn to step ii), the theoretical interpretation of the above results 
in terms
of GVD or the CDP. Both pictures have in common the basic concept of virtual
transitions
of the photon to $q\bar{q}$ (or $q\bar{q}g$) vector states with subsequent
diffractive scattering from the proton. Provided the configuration of the
$q\bar{q}$ states the photon is coupled to, and the generic structure of
two-gluon exchange \cite{11} 
in the scattering from the proton is taken into account,
(off-diagonal) GVD becomes identical \cite{8} to the CDP. While GVD is
conventionally
formulated in terms of integrals over the masses of the propagating 
($q\bar{q}$)
vector states, and the two-dimensional momentum transfer (carried by the
gluon), the CDP involves integration over the product of the square of the
photon wave function and the ($q\bar{q}$)p (`colour-dipole') cross-section
in transverse position space.

For the subsequent discussion, it will be useful, to note the 
relationship \cite{8}
between the colour-dipole cross-sections in (two-dimensional) transverse 
position space and in
momen\-tum-transfer space that guarantees the generic structure of two-gluon
exchange,
\begin{equation}
\sigma_{(q \bar q)p} (r^2_\bot , z, W^2) =
\int d^2 l_\bot \tilde\sigma_{(q \bar q)p} (l^2_\bot , z, W^2) (1 - 
\exp (-i \vec l_\bot \cdot \vec r_\bot)).  
\end{equation} 
Indeed from (4), 
\begin{equation} 
\sigma_{(q\bar q)p} (r^2_\bot , z, W^2) \longrightarrow  
\cases{ 0, & for $r_\bot \rightarrow 0$, \cr 
\int d^2 l_\bot \tilde\sigma_{(q\bar q)p} (l^2_\bot , z, W^2), & for 
$r_\bot \rightarrow \infty$,\cr}  
\end{equation} 
thus fulfilling what has been called `colour transparency' \cite{9}
and what indeed guarantees the 
generic structure of two-gluon exchange. This is explicitly seen \cite{8}, 
when representing $\sigma_{\gamma^{*} p}(W^{2}, Q^{2})$ in momentum space. 
In connection with (4) and (5), we remind the reader of the notation being  
employed, the configuration of  
$q \bar q$ states being described by the (transverse) interquark separation,
$r_\bot$, and the (light cone) variable, $z$, that is related to the 
$q \bar q$-rest-frame angle by $4 z (1-z) \equiv \sin^2 \theta$. 
 
From (5), $\tilde{\sigma}_{(q\bar{q})p}(l^{2}_\bot,z,W^{2})$ should vanish
sufficiently rapidly to yield a convergent integral. Accordingly, it may be 
suggestive to assume a Gaussian in $l^{2}_\bot$ for 
$\tilde{\sigma}_{(q\bar{q})p}(l^{2}_\bot ,z,W^{2})$. 
Actually, explicit calculations become much simpler if, without much loss of
generality,
instead of a Gaussian a $\delta$-function located at a finite value of 
$l^{2}_\bot$ 
is used \cite{8} as an effective description of 
$\tilde{\sigma}_{(q\bar{q})p}(l^{2}_\bot ,z,W^{2})$. 

Accordingly, we will adopt the simple ansatz, 
\begin{equation}
\tilde{\sigma}_{(q\bar{q})p}(l^{2}_\bot,z,W^{2})= \sigma^{(\infty)}(W^2)  
\frac{1}{\pi} \delta (l^2_\bot - z (1 - z) \Lambda^2 (W^2)). 
\end{equation} 
This ansatz associates with any given energy, $W$, an (effective) fixed value
of (two-dimensional gluon) momentum transfer, $|\vec l_\bot |$, determined 
by the so far unspecified function $\Lambda (W^2)$.
The ansatz (6) also incorporates the assumption that `aligned', 
$z\rightarrow 0$,
configurations \cite{12}
of the $(q\bar{q})$ pair absorb vanishing, $l^{2}_\bot \rightarrow 0$, 
gluon momentum. For the subsequent interpretation of our results, we note
the explicit form of  
the transverse-position-space colour-dipole cross section, obtained by
substituting
(6) into (4),
\begin{eqnarray}
& & \sigma_{(q\bar{q})p}(r^{2}_\bot,z,W^{2}) = 
\sigma^{(\infty)}(W^2)\left( 1 - J_0 \left( r_\bot \cdot \sqrt{z(1-z)}\Lambda
(W^2)\right)\right)  \\
&\cong & \sigma^{(\infty)} (W^2) \cases{ \frac{1}{4} z (1-z) \Lambda^2 (W^2)
r^2_\bot , & for $\frac{1}{4} z (1-z)\Lambda^2 (W^2) r^2_\bot \longrightarrow
0$, \cr
 & \cr
1 & for $\frac{1}{4} z (1 - z)\Lambda^2 (W^2) r^2_\bot \longrightarrow\infty$.
\cr} \nonumber
\end{eqnarray}
The limit of $\sigma^{(\infty)}(W^{2})$ in the second line of the approximate
equality in (7) actually stands for an oscillating behaviour
of the Bessel function, $J_{0}(r_\bot\sqrt{z(1-z)}\Lambda(W^{2}))$, around
$\sigma^{(\infty)}(W^{2})$, when its argument tends towards infinity.
Apart from these oscillations, the behaviour of
$\sigma_{(q\bar{q})p}(r^{2}_\bot,z,W^{2})$ in (7) is identical
to the one
obtained, if the $\delta$-function in (6) is replaced by a Gaussian. 
Concerning
the high-energy behaviour of $\sigma_{(q\bar{q})p}(r^{2}_\bot,
z,W^{2})$, we
note that it is consistent with unitarity restrictions, provided a decent
high-energy behaviour is imposed on $\sigma^{(\infty)}(W^{2})$.

We stress that the ansatz (6) is by far not as specific as it might appear 
at first sight. 
It constitutes a simple effective realization, compare (7), of the 
underlying requirements of colour transparency, (4), (5), and hadronic 
unitarity
for the colour-dipole cross section. The unitarity requirement enters via 
the decent high-energy behaviour of $\sigma^{(\infty)}(W^2)$.

Referring to \cite{10} for details, we note that the ansatz (6) allows one
to simplify the GVD expression \cite{8} for $\sigma_{\gamma^{*}p}(W^{2},
Q^{2})$ 
(by integrating over $d^{2}l$ and $dz$) to become
\begin{equation}
\sigma_{\gamma^{*}p}(W^{2},Q^{2})=
\sigma_{\gamma p} (W^2) \frac{(I_T + I_L)}{I_T |_{Q^2 = 0}} , 
\end{equation}
where $\sigma_{\gamma p}(W^{2})$ denotes the photoproduction cross section,
$\sigma_{\gamma^{*}p}(W^{2},Q^{2}=0)$, and the dimensionless quantities 
$I_{T}$
and $I_{L}$ contain integrations over the squares of the ingoing and 
outgoing masses, $M$ and 
$M^{'}$, of the $q\bar{q}$ states coupled to the ingoing and outgoing photon 
in the (virtual) forward-Compton-scattering amplitude. Expression (8) 
contains the requirement
of a smooth transition of $\sigma_{\gamma^* p} (W^2, Q^2)$ 
to photoproduction; it allowed us, to eliminate
$\sigma^{(\infty)}(W^{2})$ in terms
$\sigma_{\gamma p}(W^{2})$. Explicitly, the integrals $I_{T}$ and $I_{L}$,
that are related to the transverse and longitudinal contributions to
$\sigma_{\gamma^{*}p}(W^2, Q^2)$, respectively, are given by\footnote{In (9)
and (10), we have suppressed an additive (compensation) term that assures that
the integration over $dM^{\prime 2}$ in the off-diagonal term has the 
correct lower limit of $M^{\prime 2} \ge m^2_0$, compare ref.\cite{8}.}
\begin{eqnarray}
I_{T} \left( \frac{Q^2}{\Lambda^2 (W^2)}, \frac{m^2_0}{\Lambda^2(W^2)} 
 \right) & &\hspace*{-0.5cm} = 
{1 \over \pi} \int^\infty_{m^2_0} dM^2 \int^{(M+\Lambda (W^2))^2}_{(M-\Lambda
(W^2))^2} dM^{\prime 2}\omega (M^2, M^{\prime 2}, \Lambda^2 (W^2))\nonumber \\
& &\hspace*{-1.5cm} 
\times                               
\left[
   \frac{M^2}{(Q^2+M^2)^2}-
   \frac{M^{\prime 2}+M^2-\Lambda^2 (W^2)}
        {2(Q^2+M^2)(Q^2+M^{\prime 2})}
\right],
\end{eqnarray}
and 
\begin{eqnarray}
I_{L} \left( \frac{Q^2}{\Lambda^2(W^2)}, \frac{m^2_0}{\Lambda^2(W^2)} 
\right) & &\hspace*{-0.5cm}= 
{1 \over \pi} \int^\infty_{m^2_0} dM^2 \int^{(M+\Lambda (W^2))^2}_{(M-\Lambda
(W^2))^2} dM^{\prime 2}\nonumber \omega (M^2, M^{\prime 2}, \Lambda^2 (W^2))\\
& &\hspace*{-0.5cm} 
\times 
\left[
   \frac{Q^2}{(Q^2+M^2)^2}-
   \frac{Q^2}
        {(Q^2+M^2)(Q^2+M^{\prime 2})}
\right],
\end{eqnarray}
where the integration measure $\omega (M^2, M^{\prime 2}, \Lambda^2 (W^2))$
fulfils
\begin{equation}
\frac{1}{\pi} \int^{(M+\Lambda (W^2))^2}_{(M-\Lambda (W^2))^2} d M^{\prime 2} 
\omega (M^2, M^{\prime 2}, \Lambda^2 (W^2))= 1 . 
\end{equation}
The explicit expression for 
$\omega(M^2, M^{\prime 2}, \Lambda^2 (W^2)))$ is given in 
\cite{8}. 
It is the form
(8) to (10) of the theory that explicitly displays the structure \cite{6,7} 
of
(off-diagonal) GVD. The appearance of $\Lambda(W^{2})$ in the integration 
limits
over $dM^{'2}$ is worth noting. One expects that the effective mass range
for off-diagonal transitions, $M^{'2} \neq M^{2}$, should increase with
increasing energy, $W$, 
thus implying that $\Lambda (W^2)$ should not be constant, but should increase
with increasing energy. 

We were able to derive explicit analytic expressions \cite{10} 
for the integrals in 
(9) and (10). In the present context, we only note that, in very good 
approximation, the 
sum of $I_{T}$ and $I_{L}$ only depends on the dimensionless combination (1).
While the general explicit expressions for $I_{T}$ and $I_{L}$ are complicated,
in the most important limits, they become simple. Indeed, 
\begin{equation}
I = I_{T}+I_{L} \equiv \cases{ \log \left( \frac{\Lambda^2 (W^2)}{Q^2 + m^2_0}
\right), & for $\Lambda^2 (W^2) \gg Q^2 + m^2_0$, \cr
   & \cr
\frac{1}{2} \frac{\Lambda^2 (W^2)}{Q^2+m^2_0} , & for $\Lambda^2 (W^2) \ll Q^2
+ m^2_0$. \cr}
\end{equation}
In addition, $I_L$ vanishes for $Q^2$ towards zero. 

As the moderate rise of photoproduction, $\sigma_{\gamma p}(W^{2})$,
with energy, and the
moderate logarithmic rise of the denominator in (8) approximately cancel each
other, according to (12), we have indeed obtained 
approximate scaling of $\sigma_{\gamma^* p} (W^2, Q^2)$ in the variable
$\eta$ defined in (1), i.e. 
$\sigma_{\gamma^* p} (W^2, Q^2) \simeq \sigma_{\gamma^* p}(\eta)$.
Moreover, the theoretically expected increase of $\Lambda^2 (W^2)$ with
increasing energy coincides with the above result, (3), of the 
phenomenological analysis of the experimental data. 

The theoretical results (8) to (12) for $\sigma_{\gamma^* p}(W^2,Q^2)$ 
were obtained by incorporating the $q\bar{q}$
configuration in the virtual photon, as known from $e^{+}e^{-}$ annihilation,
as well as 
the generic structure of two-gluon exchange into the ansatz for the virtual 
Compton-forward-scattering amplitude at low $x$. 
As stressed before, the simplifying $\delta$-function ansatz (6) is to be 
seen as an effective realization of the generic two-gluon exchange structure,
combined with hadronic unitarity, without much loss of generality.

We turn to the analysis of the experimental data in terms of the 
theoretical results in (8)
to (12). This essentially amounts to introducing an empirically satisfactory
parameterization for the
photoproduction cross-section, $\sigma_{\gamma p}(W^{2})$, and to determining
the threshold mass, $m_{0}^{2}$, and the energy dependence of 
$\Lambda^{2}(W^{2})$ in fits to the
experimental data. 

Adopting a Regge parameterization for 
$\sigma_{\gamma p}(W^{2})$,
\begin{equation}
\sigma_{\gamma p}(W^{2})= A_R \cdot (W^2)^{\alpha_R -1} + A_P \cdot (W^2)^
{\alpha_P-1},
\end{equation}
where $W^2$ is to be inserted in units of GeV$^2$ and \cite{13}
\begin{eqnarray}
A_R & = & 145.0 \pm 2.0\; {\rm \mu b} , \nonumber \\
\alpha_R & = & 0.5    \\
A_P &  = & 63.5 \pm 0.9\; {\rm \mu b} , \nonumber \\
\alpha_P & = & 1.097 \pm 0.002 , \nonumber 
\end{eqnarray}
we again proceed in two steps. 

In a first step, we do not impose any specific form for the
functional dependence of $\Lambda^{2}(W^{2})$ 
except for the (technically necessary) assumption that $\Lambda^2 (W^2)$
can be represented by a piecewise
linear function of $W^2$. 
A fit to the experimental data on $\sigma_{\gamma^* p} (W^2, Q^2)$ then
determines the values of $\Lambda^2 (W^2_i)$, with $i = 1, ..., N$, that
define the piecewise linear function, $\Lambda^{2}(W^{2})$. 
In fig.2, we show the result of this procedure, $\Lambda^2 (W^2_i)$ with
$i=1, ..., 46$, including errors, obtained from the fit to the 
experimental data. 
For the fit, the restrictions of $x \le 0.1$ and $Q^2 \le 100$ GeV$^2$ 
were applied to the data.

In a second step, we adopt the
power-law ansatz (2), and again perform a fit to the data for $\sigma_
{\gamma^* p}(W^2, Q^2)$.
The agreement of the resulting curve for $\Lambda^{2}(W^{2})$ with the 
piecewise linear fit result in fig.2 shows that the power-law ansatz for
$\Lambda^{2}(W^{2})$ is borne out by the data within the
theoretical framework for $\sigma_{\gamma^{*} p}$ in (8) to (11), that 
specifies the $Q^{2}$ dependence. From the fit to the data under the 
restriction of $x\leq 0.01$ and $Q^{2}\leq 100$GeV$^{2}$, we obtained
\begin{eqnarray}
m_{0}^{2} & = & 0.16 \pm 0.01\; {\rm GeV^{2}}, \nonumber \\
C_{1}     & = & 0.34 \pm 0.05,             \\
C_{2}     & = & 0.27 \pm 0.01 , \nonumber \\
W_{0}^{2} & = & 882 \pm 246\; {\rm GeV^{2}},  \nonumber
\end{eqnarray}
with $\chi^{2}/ndf=1.15$.

The result (15), in particular the value of the exponent $C_2$ that 
determines the rise of $\Lambda^2(W^2)$ with energy, is in reasonable
agreement with the result (3) of the model-independent analysis\footnote{When
plotted, including errors, there is a significant overlap of $\Lambda^2(W^2)$
with the parameters from (3) and (15), respectively.}.
This implies that the $Q^2$ dependence of the data is correctly reproduced by
the GVD/CDP in (8) to (12). In other words, our procedure that 
combines the model-independent analysis of the data with the one based on
the GVD/CDP, has provided us with successful tests of the $W^2$- and
$Q^2$-dependence that are independent from each other. 

According to (7), the $W^2$ dependence of $\Lambda^2(W^2)$, 
displayed in fig.2, 
determines the energy dependence of the colour-dipole cross section. The 
DIS experiments at low x directly measure this quantity, in particular for
$Q^2 \ge \Lambda^2(W^2)$.

We have verified that the  replacement of the power-law ansatz (2) for
$\Lambda^2(W^2)$ by a logarithmic  one, 
\begin{equation}
\Lambda^2(W^2)=C_1^\prime \log (W^2/{W^\prime}^2_0+C_2^\prime),
\end{equation}
leads to an equally good fit to the data.  One finds
\begin{eqnarray}
m_{0}^{2} & = & 0.157 \pm 0.009\; {\rm GeV^{2}}, \nonumber \\
C_1^\prime     & = & 1.64 \pm 0.14,   \\
C_2^\prime     & = & 4.1 \pm 0.4 , \nonumber \\
{W^\prime}^2_0 & = & 1015 \pm 334\; {\rm GeV^{2}},  \nonumber
\end{eqnarray}
with $\chi^{2}/ndf=1.19$.

In fig.3, we show an explicit comparison of the experimental data with 
the GVD/CDP predictions. The (approximate) coincidence of the theoretical
predictions for various values of $W^2$ demonstrates the scaling of the theory
in terms of the low-x scaling variable $\eta$. Figure 3a, with the 
restrictions $x<0.01$ and
$Q^2<100$ GeV$^2$ imposed on the data (as in the above fit) shows the good 
agreement between theory and experiment. In fig.3b, we show the 
deviations between theory and experiment, when the data for $x\ge 0.01$ are
plotted\footnote{The fact that the model-independent analysis yields 
scaling for $x<0.1$, while fig.3b demonstrates violations for $x>0.01$
needs further investigation beyond the scope of the present note.}

Finally, fig.4a demonstrates agreement of the GVD/CDP with experiment in a 
representation of $\sigma_{\gamma^* p} (W^2, Q^2)$ against $W^2$ for fixed
values of $Q^2$. A subsample of all data used in the fit is presented for 
illustration.  

The explicit analytical form of the
theoretical expression for the cross-section, $\sigma_{\gamma^{*} p}
(W^2,Q^2)$, allows us
to investigate its behaviour at energies far beyond the ones being explored at
HERA. According to (8) with (12), at any fixed $Q^{2}$, we have a strong 
power-like increase with
energy, as $\Lambda^{2}(W^{2})$, while, finally, for sufficiently large 
energy,
the power law turns into a logarithmic rise implying an energy
behaviour as in photoproduction. This is explicitly seen in fig.4b. 
The approach to the asymptotic $W$ dependence becomes slower, if the power-law 
ansatz  for $\Lambda^2(W^2)$ in (2) is replaced by the logarithmic one in
(16).
 
The  transition from a strong power law
(or `hard') rise with energy to the soft rise in photoproduction is obviously
related to the behaviour of the dipole cross-section (7) that in turn is 
largely
dictated by the generic two-gluon exchange structure (4), (5) and the unitarity
restriction on the growth of $\sigma^{(\infty)}(W^2)$.  
At any (sufficiently small) fixed value of $r_\bot$,
(corresponding to an approximately fixed value of $Q^{2}$), the dipole
cross-section rises rapidly, as $\Lambda^{2}(W^{2})$, to finally
settle down to the limiting value of $\sigma^{(\infty)}(W^{2})$, according to 
the second line on the right-hand side in (7). As seen in fig.4b,
the scale for this transition to occur is extremely large, however,
unless $Q^{2}$ is very small.
It appears that even THERA energies of order $W^2 \cong 10^6$ GeV$^2$ may be 
too small to see this transition in the energy dependence, except at
sufficiently small $Q^2$.  

The necessary extension of the present investigation to a careful treatment 
of 
charm and of the diffractively produced final states in general is beyond the
scope\footnote{Compare, however ref.\cite{15} for a treatment of 
vector-meson production} of the present work. 

The closest in spirit to the present investigation is the work by 
Forshaw, Kerley and Shaw \cite{a}
and by Golec-Biernat and W\"usthoff \cite{b}\footnote{While the present
work was in progress we became aware of ref.\cite{c}, 
where the observation of a scaling behaviour of $\sigma_{\gamma^* p}(W^2, 
Q^2)$
within the framework of ref.\cite{b} is being reported}. 
While we agree with the general picture of low-x DIS drawn by these 
authors, there are numerous essential differences though. 
In our treatment, the dependence of the colour-dipole cross section on the
configuration variable $z$ is taken into account in contrast to 
refs.\cite{a} and \cite{b}. 
Our dipole cross section does not depend on $Q^2$, in agreement with the 
mass-dispersion relations (9), (10), but in distinction from the $Q^2$
(or rather $x$) dependence in ref.\cite{b}.
Decent high-energy behaviour at any $Q^2$ (``saturation'') follows from 
the underlying assumptions
of colour transparency (the generic two-gluon exchange structure) and 
hadronic unitarity in distinction from the two-pomeron ansatz in 
ref.\cite{a}
and in ref.\cite{d}
that needs modification at energies beyond the ones explored at HERA\footnote{
For additional references and a report on a recent discussion meeting on the 
CDP, we refer to ref.\cite{e}}.

In conclusion, a unique picture, the GVD/CDP, emerges for DIS in the 
low-x diffraction region. In terms of the (virtual) Compton-forward-scattering
amplitude, the photon virtually dissociates into $(q \bar q)$ vector states 
that propagate and undergo diffraction scattering from the proton as 
conjectured in GVD a long time ago. Our knowledge on the 
photon-$(q \bar q)$ transition from $e^+ e^-$ annihilation together with the
gluon-exchange dynamics from QCD allows for a much more detailed 
theoretical description of $\sigma_{\gamma^* p}(W^2, Q^2)$ than available at
the time when GVD was introduced. In terms of the GVD/CDP, experiments on 
DIS at low x measure the energy dependence of the $(q \bar q)$/colour-dipole 
proton cross section, $\sigma_{(q \bar q)p} (r^2_\bot , z, W^2)$. 
A strong energy dependence of
this cross section for small interquark separation 
(not entirely unexpected within the GVD/CDP) 
is extracted from the data at large $Q^2$. 
The combination of colour transparency (generic two-gluon-exchange structure)
with hadronic unitarity then implies that for any interquark separation 
the strong increase of the colour-dipole cross section, at sufficiently high energy, will settle 
down to the smooth increase of purely hadronic
interactions. As a consequence, also the strong increase with energy 
of $\sigma_{\gamma^* p}
(W^2, Q^2)$ at large $Q^{2}$ will eventually reach the behaviour observed in 
$(Q^2 = 0)$ photoproduction and hadron-hadron interactions.

\noindent\\[2ex]
{\it Acknowledgement}\hfill\\
One of us (D.S.) thanks the theory group of the Max-Planck-Institut f\"ur 
Physik in M\"unchen, where part of this work was done, for warm hospitality.
Thanks to Wolfgang Ochs for useful discussions, and
particular thanks to Leo Stodolsky for his insistence that there should 
exist a simple scaling behaviour in DIS at low x. 
We thank G. Cvetic for useful discussions and collaboration during 
the early stages of this 
work, and we also thank John Dainton and A.B. Kaidalov 
for useful discussions at `Diffraction2000' in Centraro (Sept. 2-7), 
where this work was first presented.

\pagebreak

\begin{figure}[ht]
\vspace*{-1.5cm}
\begin{center}
{\centerline{\epsfig{file=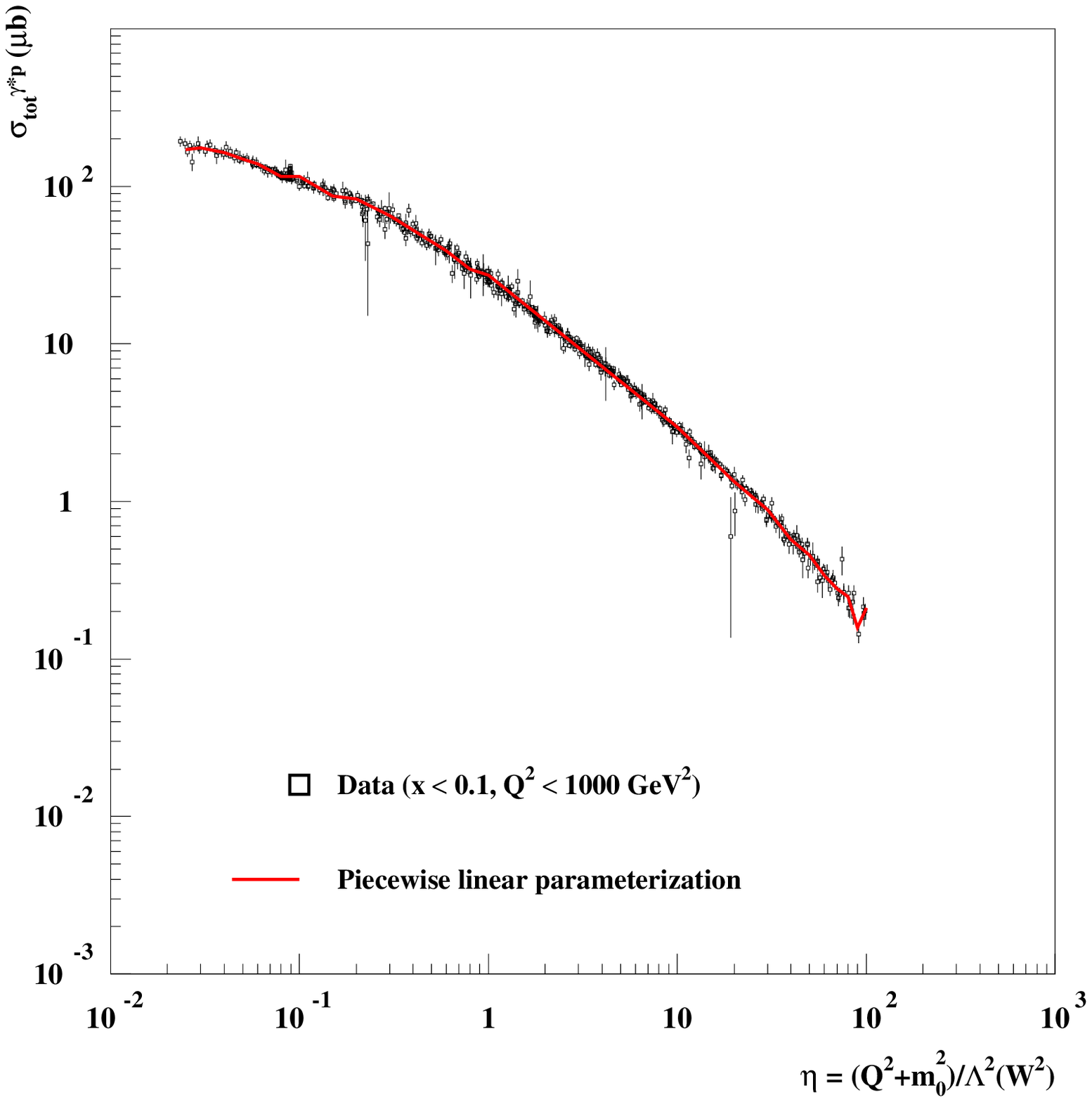,width=14.0cm}}}
\label{fig1}
\caption{The experimental data for $\sigma_{\gamma^* p} (W^2,Q^2)$ for 
$x \simeq Q^2/W^2 < 0.1$ vs. the low-x scaling variable 
$\eta = (Q^2 + m^2_0) / \Lambda^2 (W^2)$.}
\end{center}
\end{figure}

\pagebreak

\begin{figure}[ht]
\vspace*{-0.5cm}
\begin{center}
{\centerline{\epsfig{file=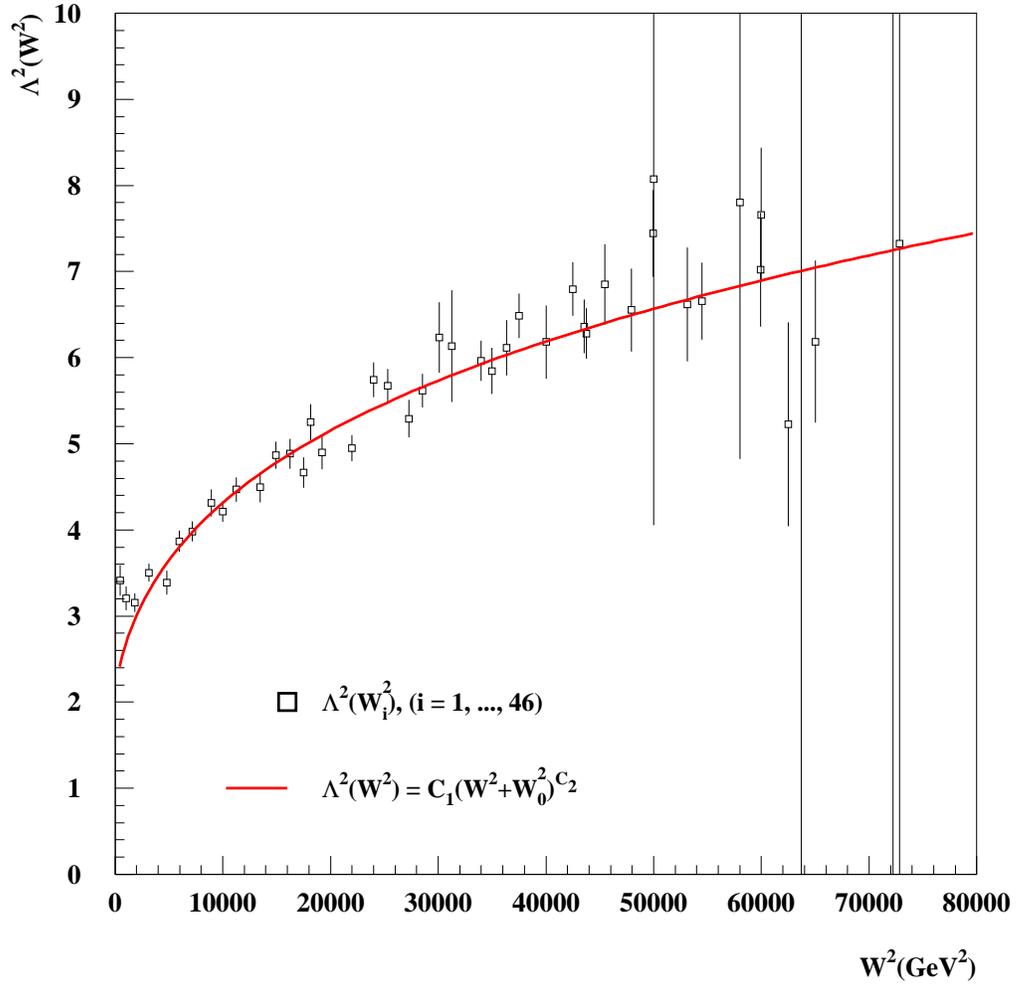,width=14.0cm}}}
\label{fig2}
\caption{The dependence of $\Lambda^2$ on $W^2$, as determined by 
a fit of the GVD/CDP 
predictions for $\sigma_{\gamma^* p}$ to the experimental data.}
\end{center}
\end{figure}

\pagebreak

\begin{figure}[ht]
\vspace*{-1.5cm}
\begin{center}
{\centerline{\epsfig{file=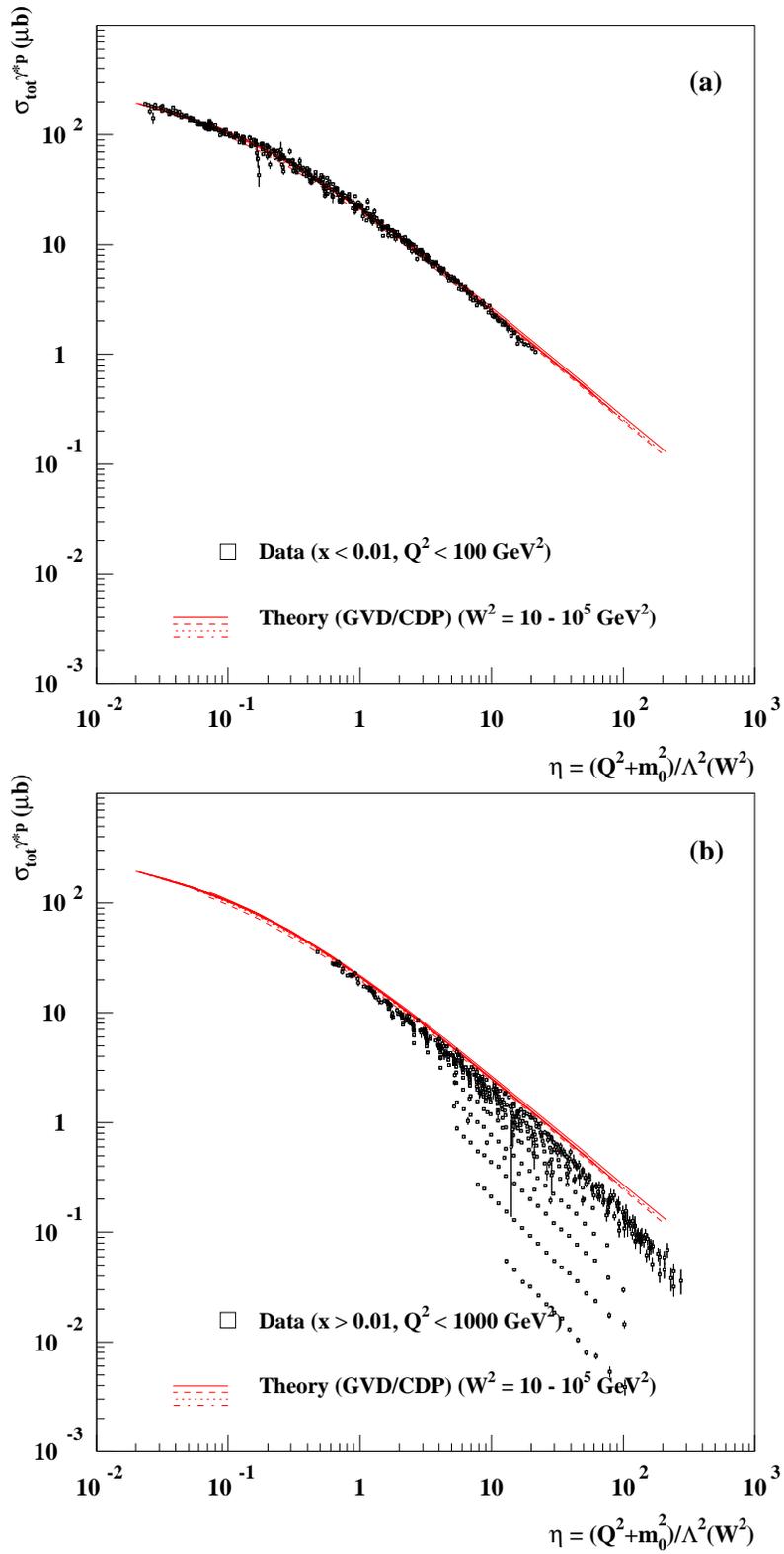,width=11.0cm}}}
\label{fig3}
\caption{The GVD/CDP scaling curve for $\sigma_{\gamma^* p}$ compared with the 
experimental data a) for $x < 0.01$, b) for $x > 0.01$.}
\end{center}
\end{figure}

\newpage

\begin{figure}[ht]
\vspace*{-1.5cm}
\begin{center}
{\centerline{\epsfig{file=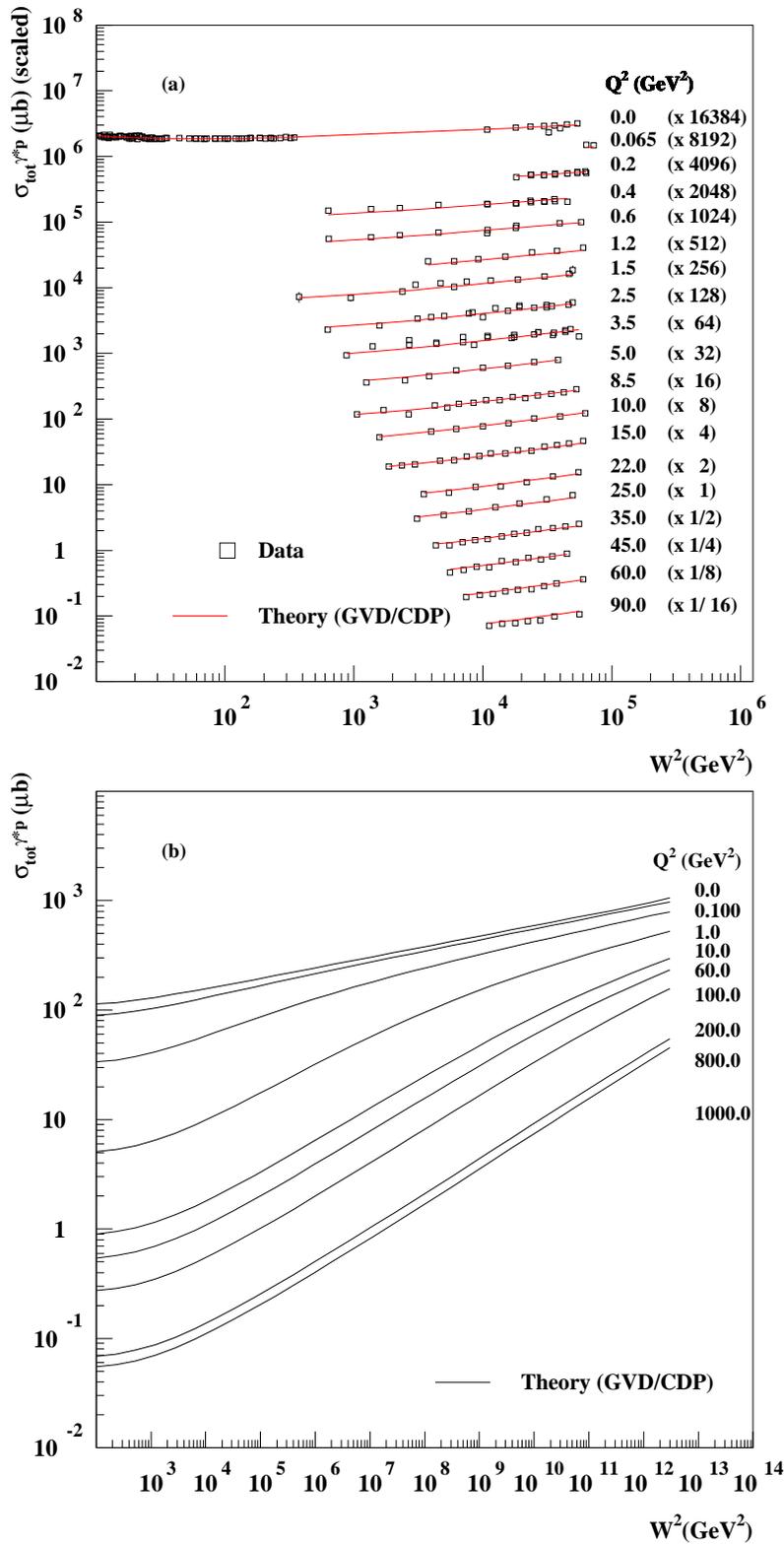,width=11.0cm}}}
\label{fig4}
\caption{The GVD/CDP predictions for $\sigma_{\gamma^* p}(W^2, Q^2)$ vs. $W^2$ 
at fixed $Q^2$\  a) in the presently accessible energy range
compared with experimental data for $x \le 0.01$, 
\  b) for asymptotic energies. The cross-section results for $Q^2=0.0
\ \mbox{GeV}^2$ refer to the total photoproduction measurements
\protect\cite{5a} and the extrapolated results at low $Q^2$
\protect\cite{13}. They were not explicitly included in the fit, but
rather the Regge parameterization (13).
}
\end{center}
\end{figure}
\pagestyle{empty}

\end{document}